\documentclass[aps,twocolumn,showpacs,preprintnumbers,amsmath,amssymb,prl]{revtex4-1}
\usepackage{epsf,amsmath,amssymb,amsfonts,verbatim,color,multirow,pifont}
\usepackage{graphicx}

\begin{document}
\title{Geometry-induced rectification for an active object}
\author{Jae Sung Lee$^{1}$} \email{jslee@kias.re.kr}
\author{Jong-Min Park$^{1}$}
\author{Jae Dong Noh$^{2}$} 
\author{Hyunggyu~Park$^{1}$} \email{hgpark@kias.re.kr}
\affiliation{{$^1$School of Physics, Korea Institute for Advanced Study, Seoul 02455, Korea}}
\affiliation{{$^2$ Department of Physics, University of Seoul, Seoul 02504, Korea}}

\newcommand{\revise}[1]{{\color{red}#1}}

\date{\today}

\begin{abstract}
Study on a rectified current induced by active particles has received a great attention due to its possible application to a microscopic motor in biological environments. Insertion of an {\em asymmetric} passive object amid many active particles has been regarded as an essential ingredient for generating such a rectified motion. Here, we report that the reverse situation is also possible, where the motion of an active object can be rectified by its geometric asymmetry amid many passive particles. This may describe an unidirectional motion of polar biological agents with asymmetric shape. We also find a weak but less diffusive rectified motion in a {\em  passive} mode without energy pump-in. This ``moving by dissipation'' mechanism could be used as a design principle for developing more reliable microscopic motors.
\end{abstract}


\maketitle

\emph{Introduction} -- Most biological systems are \emph{active} in that they are self-propelled, i.e., driven by mechanical forces generated via an internal mechanism consuming chemical fuels~\cite{active_review}. This is in marked contrast to \emph{passive} systems such as a Brownian particle of which stochastic motions are governed by external reservoirs. The activeness leads to various unique features clearly distinguished from passive systems. For example, active systems show time-scale-dependent diffusivity due to colored noise~\cite{active_exp1,active_exp2,active_exp3,active_exp4,active_exp5}, aggregation due to repulsive force~\cite{active_aggre}, efficiency enhancement~\cite{active_engine1, active_engine2, active_engine3}, and unconventional entropy production~\cite{active_engine3, Kwon, Lee_old, active_entropy1, active_entropy3}.

In many literatures, this self-propelled motion has been encoded into velocity-dependent driving forces fueling the active object  such as molecular motors, moving cells, and bacteria at the phenomenological level~\cite{active_review}. For example, typical driving forces  are given in the Rayleigh-Helmholtz (RH)~\cite{RH1, RH2, RH3}, the depot~\cite{D1, D2, D3, D4}, and  the Schienbein-Gruler (SG) models~\cite{SG0, SG1} as
\begin{align}
\textbf{\textit{F}}_\textrm{drv} = -\Gamma(\textbf{\textit{v}})\textbf{\textit{v}},~\textrm{where }\Gamma(\textbf{\textit{v}}) = \left\{
\begin{array}{ll}
 \hat{\gamma} +\omega \textbf{\textit{v}}^2 &: \textrm{RH} \\
 \frac{\hat{\gamma}}{1+\zeta \textbf{\textit{v}}^2 }  & : \textrm{depot}\\
\hat{\gamma}v_0/|\textbf{\textit{v}}|  & : \textrm{SG}
\end{array}
\right. , \label{eq:threeforces}
\end{align}
where $\textbf{\textit{v}}$ is the velocity of an active particle, $v_0$ is a velocity unit defined in Eq.~\eqref{eq:rescaling}, and $\hat{\gamma}$, $\omega$, and $\zeta$ are tunable parameters.
Here, we take $\omega, \zeta >0$ for the stability.

\begin{figure*}
\centering
\includegraphics[width=0.95\textwidth]{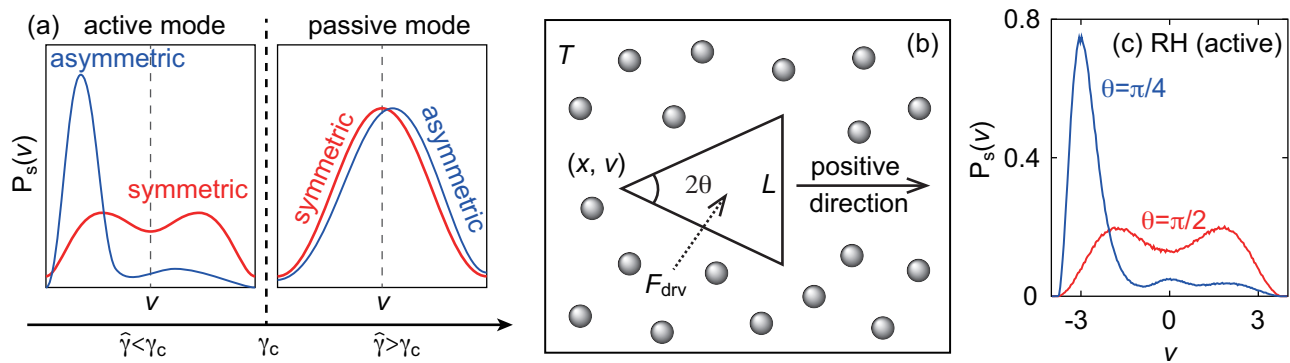}
\caption{(a) Schematic of the active and the passive mode in one dimension. When $\hat{\gamma} < \gamma_\textrm{c}$ ($\hat{\gamma} > \gamma_\textrm{c}$) with a symmetric particle shape, the velocity distribution becomes symmetric bimodal (unimodal), which indicates the active (passive) mode as the left (right) red curve. When the shape of particle becomes asymmetric, one peak is amplified over the other for the active mode, while the peak position moves slightly aside for the passive mode (see the blue curves).  (b) Schematic of the model. The t-particle is immersed in a reservoir with temperature $T$
and moves along the horizontal direction (one-dimensional motion). The reservoir consists of $N$ identical r-particles.
The driving force, $F_\textrm{drv}$, is applied to the t-particle.
(c) The steady-state velocity distributions for the RH model in the active mode. Red and blue curves are plots of numerical data for $\theta=\pi/2$ (symmetric) and $\pi/4$ (asymmetric), respectively. See SM Fig.~S1 for numerical details and
also for the distributions for the depot and the SG models.
 } \label{fig:schematic}
\end{figure*}

The active motion (or active mode) emerges with a nonzero self-propelling velocity
for $\hat{\gamma} <\gamma_\textrm{c}$, which is represented by non-zero peak positions in
the velocity distribution (see the left red curve in Fig.~\ref{fig:schematic}(a)).
Specifically, $\gamma_\textrm{c}= -\gamma $ for the RH and the depot model, and $\gamma_\textrm{c}= 0$ for the SG model with the friction coefficient $\gamma$ for the active particle moving in a passive reservoir.
For $\hat{\gamma}>\gamma_\textrm{c}$, the velocity distribution becomes unimodal with no finite self-propelling velocity
(see the right red curve in Fig.~\ref{fig:schematic}(a)), which indicates the passive  mode.

Even in the active mode, a self-propelled particle does not prefer any particular direction like in a standard run-and-tumble motion~\cite{active_review}, thus there exists no rectified motion (no net particle current) in the long-time limit. The zero current is a natural consequence originating from the symmetry of the driving forces  with no favored direction in Eq.~\eqref{eq:threeforces}. In order to observe a nonzero current, one may resort to a collective motion of interacting self-propelled particles, which is not of our interest here, because it is not applicable to a microscopic engine rectifying a single object motion.

In the meantime, the rectification of thermal fluctuations in a Brownian motor has been already reported at the level of nonlinear response~\cite{Broeck, Meurs}. The essential ingredient for the nonzero current is the interplay of a nonequilibrium situation (temperature gradient) and a geometric asymmetry of the motor. Interestingly, a `passive' ratchet surrounded by many `active' bacteria
was found to exhibit a persistent rotational motion (rectified current) in recent experiments~\cite{active_current5, active_current6} and numerical simulations~\cite{active_current4}. This triggered a flurry of subsequent research~\cite{active_current1, active_current2, active_current3, active_current7, active_current8}, due to a realistic applicability for designing a microscopic motor in biological environments. Note that this situation also combines nonequilibrium-ness (activity) and spatial asymmetry of the ratchet.

In this Letter, we introduce the reverse situation where an asymmetric active particle is immersed in a reservoir of passive particles.
This may describe the motion of a {\em polar} biological agent or motor inside a passive fluid such as water~\cite{active_review}. We find that the geometric asymmetry plays a crucial role for the rectified motion along with nonequilibrium-ness caused by the driving force in Eq.~\eqref{eq:threeforces}. We derive analytically the explicit formula for the rectified current  and compare with numerical results via extensive molecular dynamics simulations. Surprisingly, a rectified current exists even in the passive mode, in particular for $\hat{\gamma}>0$ where the energy only dissipates (no energy pump-in).  This is possible because thermal fluctuations are significantly reduced in this case, compensating the energy needed for generating an average current. This is why we coin the term ``moving by dissipation'' for this rectifying mechanism. In practice, this type of
rectification provides a weak but reliable (less diffusive) directed motion even in highly fluctuating thermal environments.

\emph{Model} -- Consider a triangular shaped particle (t-particle) with mass
$M$ and vertical cross section $L$ immersed in a reservoir with temperature
$T$ as illustrated in Fig.~\ref{fig:schematic}(b).
For simplicity, we constrain the t-particle motion along the horizontal direction only. Its apex angle,
position, and velocity are denoted as $2 \theta$, $x$, and $v$,
respectively. The reservoir consists of $N$ identical circular shaped
particles (r-particle) with mass $m$ and radius $R$ which move inside a two-dimensional
square box with side length $l$ with periodic boundary conditions.
We assume that $m\ll M$ and $R\ll L\ll l$.

The stochastic motion of the t-particle is induced by elastic collisions with r-particles, which are
equipped with the Langevin thermostat. Details of collision dynamics and numerical simulations~\cite{Broeck, additivity}
are described in Supplementary Material (SM) I.
Without any driving force, the t-particle reaches a thermal equilibrium
 with zero mean velocity. With a one-dimensional horizontal force in Eq.~\eqref{eq:threeforces} applied on the t-particle, the
 system can be driven out of equilibrium. The red and blue curves in Fig.~\ref{fig:schematic} (c) represent the numerical data for the steady-state velocity distributions of the t-particle with the symmetric shape $(\theta=\pi/2)$ and the asymmetric shape $(\theta=\pi/4)$ in the active mode for the RH model, respectively.
Similar distributions for the depot and the SG models are presented in SM Fig.~S1.
Clear two symmetric red peaks denote the emergence of the nonzero self-propelled velocity with the zero average current
for the symmetric case. In contrast, the shape asymmetry breaks the symmetry of the peaks, thus the active motion is
rectified with a nonzero average current. Typical `run-and-tumble' trajectories for the symmetric and asymmetric particle
are shown in SM Fig.~S2.

\begin{figure*}
\centering
\includegraphics[width=0.95\linewidth]{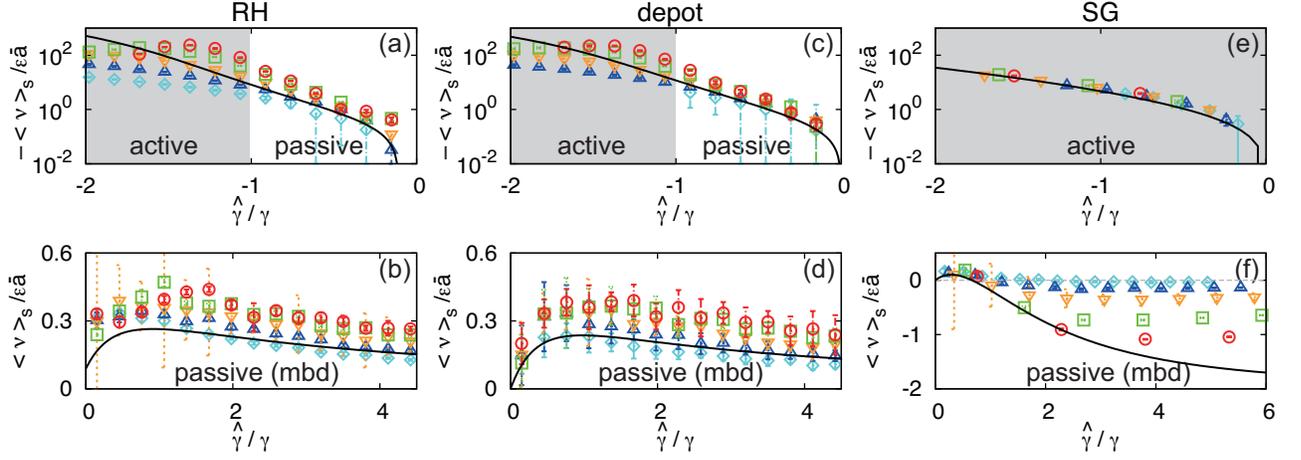}
\caption{ $\hat{\gamma}$ dependence of the rectified velocity for various self-propelling mechanisms. (a), (c), and (e) are the scaled steady-state velocities $\langle \nu \rangle_\textrm{s} / (\bar{a} \epsilon)$ as a function of $\hat{\gamma}/\gamma$ for the RH, the depot, and the SG models for negative $\hat{\gamma}$, respectively.
(b), (d), and (f) are for positive $\hat{\gamma}$. Solid curves are analytic results for respective mechanisms. Cyan $\diamond$, blue $\vartriangle$, gold $\triangledown$,  green $\square$, red $\circ$ points denote data for $M=5$, $10$, $20$, $50$, and $100$, respectively, with $m=1$ fixed. `mbd' stands for `moving by dissipation'. } \label{fig:othermodel}
\end{figure*}

In the passive mode, we find a single peak at the zero velocity for the symmetric case, but at a small non-zero velocity for the asymmetric case, representing a {\em weak} rectified current, compared to that in the active mode. Figure~\ref{fig:othermodel} shows the
$\hat{\gamma}$ dependence of the average current obtained from simulations and also from the analytic derivation in the
small $m/M$ limit shown below.

\emph{Perturbative analysis of the model} -- To understand the emergence of a nonzero current analytically, we present a perturbation theory for the t-particle dynamics for small $m/M$, which is reasonable in realistic situations. We also assume that r-particles are always in equilibrium. Then, our model can be described by the kinetic theory introduced in Ref.~\cite{Broeck, Meurs} with the addition of an external driving force.

The probability density of the t-particle velocity $v$ at time $t$, $P(v,t)$, can be described by the following Boltzmann-Master equation:
\begin{align}
\frac{\partial P(v,t)}{\partial t} = (\mathcal{L}_\textrm{res} +  \mathcal{L}_\textrm{drv} ) P(v,t), \label{eq:Boltzmann_Master}
\end{align}
where $\mathcal{L}_\textrm{res}$ and $\mathcal{L}_\textrm{drv}$ are
operators  representing the effects of the reservoir and the driving
force, respectively. More specifically, $\mathcal{L}_\textrm{res}$ can be written as
\begin{align}
\mathcal{L}_\textrm{res}=\sum_{n=1}^{\infty}  \frac{(-1)^n}{n!} \frac{\partial^n}{\partial v^n} a_n (v) ,
\label{eq:reservoir_operator}
\end{align}
where a Kramers-Moyal coefficient $a_n(v)$ is defined as $
a_n (v) \equiv \int r^n W(v+r|v) dr$ with the transition rate $W(v'|v)$  from $v$ to $v'$ induced by elastic collisions with r-particles~\cite{Meurs, Risken, vanKampen}. The explicit calculation of $a_n(v)$
is presented in SM II.  $\mathcal{L}_\textrm{drv}$ is given by
\begin{align}
\mathcal{L}_\textrm{dis}=   -\frac{\partial}{\partial v} \frac{F_\textrm{drv} }{M},
\label{eq:dissipation_operator}
\end{align}
with $F_\textrm{drv}=-\Gamma(v) v$ in Eq.~\eqref{eq:threeforces}.

It is convenient to introduce dimensionless variables
\begin{align}
\nu \equiv v / v_0~~\left( v_0\equiv \sqrt{ {k_\textrm{B} T}/{M} }\right) ~~\textrm{ and }~~ \tau= \gamma t/M,   \label{eq:rescaling}
\end{align}
where
the t-particle friction coefficient $\gamma$ is obtained as
\begin{align}
\gamma \equiv 4 L \rho \sqrt{\frac{m k_\textrm{B} T }{2\pi} } (1+\sin \theta)~,  \label{eq:gamma}
\end{align}
with the r-particle density $\rho=N/l^2$ (see SM II).
Then, Eq.~\eqref{eq:Boltzmann_Master} is rewritten as
\begin{align}
    \frac{\partial P(\nu,\tau)}{\partial \tau} =
     \sum_{n=1}^{\infty} \frac{(-1)^n}{n!}  \frac{\partial^n}{\partial \nu^n} A_n (\nu) P(\nu,\tau), \label{eq:rescaledFP}
\end{align}
where $A_n (\nu)$ is the modified  Kramers-Moyal coefficient defined as
\begin{align}
A_n(\nu) = \frac{M}{\gamma  v_0^{n}}  a_n ( v_0 \nu )
-   G ( \nu ) \nu \delta_{n,1}
\end{align}
with $P( \nu,\tau ) = v_0 P ( v, t)$
and $G ( \nu ) = \Gamma ( v_0 \nu )/\gamma$.

We perform a perturbation expansion with the small parameter
$\epsilon \equiv \sqrt{m/M}$.
Up to ${\cal O}(\epsilon)$, we find in SM II
\begin{align}
 A_1 (\nu) &\approx   - \left[1 + G(\nu) \right] \nu + \bar{a} \left( 1 - \nu^2 \right) \epsilon, \nonumber \\
 A_2 (\nu) &\approx  2 + 6 \bar{a} \nu \epsilon , \nonumber  \\
 A_3 (\nu) &\approx   - 12 \bar{a} \epsilon , \nonumber \\ 
 A_n (\nu) &= {\cal O} (\epsilon^2), ~~\textrm{for}~ n \geq  4,
  \label{eq:approxAn}
\end{align}
where the asymmetric factor  $\bar{a} = \frac{\sqrt{2 \pi}}{4} ( 1 - \sin \theta )\ge 0.$
Using Eqs.~\eqref{eq:rescaledFP}, \eqref{eq:approxAn} and
the expansion of $P(\nu,\tau) \approx P^{(0)}(\nu,\tau)+\epsilon P^{(1)}(\nu,\tau)$, we can set up the equations as follows:
\begin{align}
\partial_\tau P^{(0)} (\nu,\tau) &=\mathcal{L}_0 P^{(0)} (\nu,\tau)  \nonumber \\
\partial_\tau P^{(1)} (\nu,\tau) & = \mathcal{L}_1 P^{(0)} (\nu,\tau) + \mathcal{L}_0 P^{(1)} (\nu,\tau)
\end{align}
where $\mathcal{L}_0$ and $\mathcal{L}_1$ are given by
\begin{align}
    \mathcal{L}_0 &= - \frac{\partial}{\partial \nu} \left[ - \left ( 1 + G (\nu) \right ) \nu - \frac{\partial}{\partial \nu} \right] \nonumber \\
    \mathcal{L}_1 &= - \bar{a} \frac{\partial}{\partial \nu} \left[ ( 1 - \nu^2 ) - 3 \frac{\partial}{\partial \nu} \nu - 2 \frac{\partial^2}{\partial \nu^2}  \right].
    \label{eq:perturbed_eq}
\end{align}
The steady-state distribution of the zeroth order is then
\begin{align}
    P_\textrm{s}^{(0)} (\nu) = \frac{1}{\mathcal{N}} e^{- \int^\nu d s  [ 1 + G (s) ]s } \label{eq:zeroPDF}
\end{align}
with the normalization factor $\mathcal{N}$. The next order is obtained by solving $\mathcal{L}_0 P_\textrm{s}^{(1)}(\nu) = - \mathcal{L}_1 P_\textrm{s}^{(0)} (\nu)$. A straightforward analysis yields
\begin{align}
&P_\textrm{s}^{(1)} (\nu) = \bar{a} g(\nu) P_\textrm{s}^{(0)} (\nu) \quad \textrm{with}\nonumber\\
  &g(\nu) =
    2 G (\nu) \nu - \int_0^\nu d s~
    G (s) \left[ 1 + 2 G ( s ) \right]  s^2 . \label{eq:g_nu}
\end{align}
Note that $g(\nu)$ is an odd function of $\nu$, i.e., $g(-\nu) = -g(\nu)$,
as $G(\nu)$ is even in $\nu$ in all models considered here. This is also
consistent with the normalization condition
$\int_{-\infty}^{\infty} d s P_\textrm{s}^{(1)} (s) = 0$.

The steady-state average of the $n$-th moment of the velocity is then
obtained up to ${\cal O}(\epsilon)$ as
\begin{align}
	\langle \nu^n \rangle_\textrm{s}
    &\approx \int_{-\infty}^{\infty} d\nu ~\nu^n \left[ 1 + \bar{a} \epsilon g(\nu)  \right]
    P_\textrm{s}^{(0)} (\nu) \nonumber \\
    &= \left\{ \begin{array}{ll}
    \bar{a} \epsilon \langle \nu^n g(\nu) \rangle_0  & (n: \textrm{odd}) \\
    \langle \nu^n \rangle_0 & (n: \textrm{even})
    \end{array}  \right. , \label{eq:vn}
\end{align}
where $\langle \cdots \rangle_0$ stands for the average over $P_\textrm{s}^{(0)} (\nu)$.
Note that the average velocity as well as its all odd moments
are ${\cal O}(\epsilon)$ and vanishes for the symmetric case ($\bar{a}=0$).
Thus the time-reversal symmetry breaking (rectified current) occurs only
with a shape asymmetry and a finite mass ratio.
All even moments responsible for the stochasticity are always
${\cal O}(1)$, the same as that for $\bar{a}=0$. The standard fluctuation is
simply given by the second moment;
$\langle (\Delta \nu)^2\rangle_s=\langle \nu^2\rangle_s -\langle \nu\rangle_s^2 \approx \langle \nu^2\rangle_0$
up to ${\cal O}(\epsilon)$.
Similar results can be derived for the t-particle with an arbitrary convex shape.

\emph{Examples} --
We first consider a simple example with $F_\textrm{drv} = -\hat{\gamma} v$, yielding
$G(\nu)=\hat\gamma/\gamma$. This model is known to describe a ``cold damping'' problem applicable to
a molecular refrigerator~\cite{Pinard, Jourdan, Kim}.
With Eqs.~\eqref{eq:zeroPDF}, ~\eqref{eq:g_nu} and \eqref{eq:vn}, we easily find the scaled {rectified} velocity  and its fluctuation as
\begin{align}
\langle \nu \rangle_\textrm{s}^\textrm{sim}  = \bar{a} \varepsilon  \frac{\hat{\gamma} /\gamma}{ (1 +\hat{\gamma}/\gamma)^2 },
\quad \langle \nu^2 \rangle_\textrm{s}^\textrm{sim} = \frac{1}{ 1 + \hat{\gamma}/\gamma }
\label{eq:ssvelocity}
\end{align}
with
$P_\textrm{s}^{(0)} (\nu) \sim \exp[-(1+\hat\gamma/\gamma) \nu^2/2]$
 (see SM III).
As expected, the rectification is due to the interplay of nonequilibrium ($\hat\gamma$) and spatial asymmetry ($\bar{a}$),
even though the driving force does not favor any particular spatial direction.
In this simple model, only the passive mode ($\hat{\gamma} > -\gamma$) is allowed due to the dynamic instability in the active region.
For $\hat\gamma>0$, we get a nonzero current without energy input by the driving force (rather dissipation only)
- ``moving by dissipation''.
This can be understood that the driving force can generate the rectification energy by
cooling down the t-particle (less fluctuation in Eq.~\eqref{eq:ssvelocity}).
Note that the t-particle can move in either direction, depending on the sign of $\hat\gamma$ and
slows down for large positive $\hat\gamma$ until the motion is fully stalled at $\hat\gamma=\infty$.

Now, we consider the three models for active dynamics as in Eq.~\eqref{eq:threeforces}.
We calculated the steady-state velocities $\langle \nu \rangle_\textrm{s}^\textrm{RH}$, $\langle \nu \rangle_\textrm{s}^\textrm{dpt}$, $\langle \nu \rangle_\textrm{s}^\textrm{SG}$, and their fluctuations for the RH, the depot,
and the SG model, respectively.
The calculation results are rather complicated, which are presented in SM III.
The velocities are plotted as solid curves in Fig.~\ref{fig:othermodel}, which show the stabilized active mode ($\hat\gamma/\gamma<-1$ for the RH and the depot model, and
$\hat\gamma<0$
for the SG model)
with a much bigger negative current, compared to that in the passive mode.
In fact, $\langle \nu \rangle_\textrm{s}$ grows indefinitely as $\hat\gamma\rightarrow -\infty$ with  huge energy pump-in by the driving force, and its fluctuation also diverges.  In the passive mode, we find a weaker positive current for the RH and the depot model, similar to the above simple example for $\hat\gamma>0$. In contrast, the SG model shows an interesting crossover behavior from a positive to negative current for positive $\hat\gamma$
with $\langle \nu \rangle_\textrm{s}^\textrm{SG}=-2 \bar{a}\epsilon$ in the $\hat\gamma\rightarrow\infty$ limit.
This may be related to a faster decay of fluctuations in the SG model: $\langle \nu^2 \rangle_\textrm{s}\sim \hat\gamma^{-2}$ (SG)
and $\sim \hat\gamma^{-1}$ (RH and depot), see SM III and SM Fig.~S7. Thus, the SG type would be more appropriate for designing reliable microscopic motors with both directional currents possible.

\emph{Numerical simulations} --
To confirm the validity of our analysis, we performed extensive molecular dynamics simulations explained in SM I.
Data points in Fig.~\ref{fig:othermodel} are the simulation results for  the rectified velocities of the asymmetric t-particle
with various values of $\hat{\gamma}$ and t-particle mass
$M=5, 10, 20, 50, 100$. All other parameters are fixed (see the caption of SM Fig.~S1).

We find that their overall behaviors agree  well with the theoretical predictions qualitatively, but with $\sim 40\%$ overestimates for the simple (SM Fig.~S4), the RH, and the depot models in the small $\epsilon$ limit. The origin of this difference is unclear yet, but probably
due to a reservoir finite-size effect, also noticed by previous studies in similar systems~\cite{Broeck, Meurs}.
For the SG model, the numerical over-estimate is much smaller, but the convergence to the small $\epsilon$ limit is
quite slow for large positive $\hat{\gamma}/\gamma$. This slow convergence is also found in
the active phase for the RH and the depot model. In order to understand the discrepancy between numerical and theoretical results
systematically, further extensive simulations are necessary, which is out of  scope of our research here.

\begin{figure}
\centering
\includegraphics[width=0.9\linewidth]{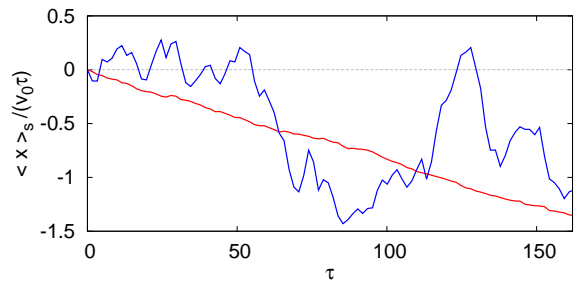}
\caption{ Trajectories of the t-particle averaged over 100 realizations driven by a constant force (blue) $f =5.1  \times 10^{-3} \gamma v_0$ and the SG force with $\hat\gamma/\gamma=9.88$ (red).
} \label{fig:vel_variance}
\end{figure}

The fluctuations are also plotted in SM Fig.~S7. They agree very well with the theoretical predictions, except
for the large-$\hat\gamma/\gamma$ region of the SG model, showing again a similar slow convergence.
For highlighting the usefulness of the moving-by-dissipation mechanism, we compare two simulated trajectories of the same t-particle driven by either constant force ($F_\textrm{drv} = f$) or the SG force, resulting in the same average velocity.
The blue (red) curve in Fig.~\ref{fig:vel_variance} is the averaged trajectory over $100$ realizations when the t-particle is driven by the constant (SG) force. This clearly shows that the moving-by-dissipation mechanism can be utilized as a motor mechanism when an accurate motion is required in a highly fluctuating environment.

\emph{Conclusion} -- Our study clearly demonstrates that a self-propelled motion can be rectified by a geometric asymmetry of the active particle shape. We also show that this rectification is possible even in the passive mode. {Especially, for $\hat{\gamma}>0$, the motion is driven by} the moving-by-dissipation mechanism which can provide a novel
design principle for developing more reliable microscopic motors. Our results are analytically derived by a relevant kinetic theory and supported qualitatively  by numerical simulations.
It is also imaginable that some microorganisms (and also nanomachines driven by chemical fuels) make use of this rectification mechanism by changing their shape asymmetrically to move in an intended direction.

\begin{acknowledgments}
Authors acknowlege the Korea Institute for Advanced Study for providing computing resources (KIAS Center for Advanced Computation Linux Cluster System). This research was supported by the NRF Grant No. 2017R1D1A1B06035497 (HP) and 2019R1A2C1009628 (JDN), and the KIAS individual Grants No. PG013604 (HP), PG074001 (JMP), PG064901 (JSL) at Korea Institute for Advanced Study.
\end{acknowledgments}

\vfil\eject

\end{document}


\title{Supplementary Material for \\
``Geometry-induced rectification for an active object''}
\author{Jae Sung Lee$^{1}$} \email{jslee@kias.re.kr}
\author{Jong-Min Park$^{1}$}
\author{Jae Dong Noh$^{2}$} 
\author{Hyunggyu~Park$^{1}$} \email{hgpark@kias.re.kr}
\affiliation{{$^1$School of Physics, Korea Institute for Advanced Study, Seoul 02455, Korea}}
\affiliation{{$^2$ Department of Physics, University of Seoul, Seoul 02504, Korea}}

\newcommand{\red}[1]{{\color{red}#1}}


\maketitle

\section{Molecular dynamics simulations}

We consider a triangular shaped particle (t-particle) with mass
$M$ in a reservoir with $N$ identical circular reservoir particles (r-particles) as shown in
Fig.~1 (b) of the main text. The t-particle moves only in the horizontal direction, while
the r-particles move inside the two-dimensional box with periodic boundary conditions.
Interactions between the r-particles are modeled as hard-disk elastic
collisions.
For convenience, an elastic collision between a r-particle and the t-particle is assumed to occur
when the r-particle center touches the boundary of the t-particle~\cite{Broeck, additivity}.

We adopt the Langevin thermostat to maintain the reservoir temperature $T$~\cite{Grest}.
Thus, the equation of motion of the $i$-th r-particle ($i=1,\cdots,N$) is given by~\cite{additivity}
\begin{align}
\textbf{\textit{v}}_i = \dot{\textit{\textbf{x}}}_i,~~m \dot{\textbf{\textit{v}}}_i = \textbf{\textit{F}}^\textrm{col}_i -\gamma_\textrm{th} \textbf{\textit{v}}_i + \boldsymbol{\xi}_i, \label{eq:thermostat}
\end{align}
where $\textbf{\textit{x}}_i$ and $\textbf{\textit{v}}_i$ are the
two-dimensional position and velocity vectors of the $i$-th r-particle,
respectively, $\gamma_\textrm{th}$ is the dissipation coefficient for the
thermostat, and $\textbf{\textit{F}}^\textrm{col}_i$ describes the interaction
force induced by collisions with the t-particle and other r-particles. $ \boldsymbol{\xi}_i$ is
the Gaussian white noise vector satisfying $\langle  \boldsymbol{\xi}_i (t)
\boldsymbol{\xi}_j^\intercal (t^\prime) \rangle =2 \gamma_\textrm{th}
k_\textrm{B} T \delta_{ij} \delta(t-t') \mathbb{I} $, where $\mathbb{I}$ is
the $N\times N$ identity matrix and $k_\textrm{B}$ is the Boltzmann constant.

The t-particle motion is affected by the elastic collisions with r-particles and also by
the driving force in Eq.~(1) of the main text.
As we constrain the t-particle motion in one dimension (horizontal direction), the equation of motion of the t-particle is
\begin{align}
v=\dot{x},~~M\dot{v}=-\textbf{\textit{F}}^\textrm{col}_x +F_\textrm{drv}, \label{eq:t-particle}
\end{align}
where $x$ and $v$ are the position and the velocity of the t-particle, $\textbf{\textit{F}}^\textrm{col}_x$ is the horizontal component of the total collision force ($\textbf{\textit{F}}^\textrm{col}=\sum_i \textbf{\textit{F}}^\textrm{col}_i$), and
the one-dimensional driving force $F_\textrm{drv}=-\Gamma(v) v$.

For numerical simulations, we set the parameter values as $l=300$, $N=200$, $k_\textrm{B} T=4$, $R=0.5$, $L=4$, $\gamma_\textrm{th}=1$, $m=1$,
 $\omega = 0.01$, and $\zeta=0.2$. The apex angle $\theta=\pi/2$ is taken for the symmetric shape (vertical rod) and
 $\theta=\pi/4$ for a typical asymmetric shape. 
  We vary $\hat{\gamma}$ and $M$ to investigate the dependence on the driving force strength ($\hat\gamma$)
  and also on the mass ratio ($m/M$). By solving the equations of motion in Eqs.~\eqref{eq:thermostat} and \eqref{eq:t-particle}
  numerically with kinematic considerations for all involved collisions, we generate numerical data for the t-particle trajectories
 in the phase space of $(x,v)$. After discarding initial transients, we calculate the steady-state velocity distribution
 $P_\textrm{s}(v)$ as well as its first and second moments, i.e.~$\langle v\rangle_\textrm{s}$ and $\langle v^2\rangle_\textrm{s}$.

\begin{figure}
	\centering
	\includegraphics[width=0.5\linewidth]{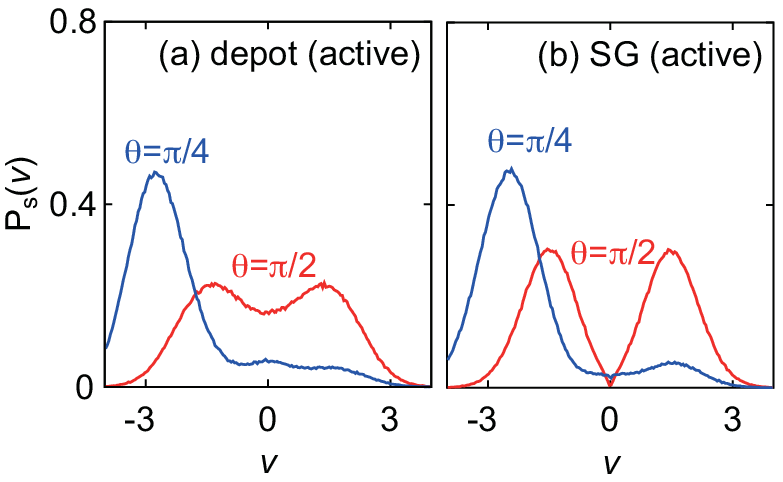}
	\caption{The steady-state velocity distributions for (a) the depot and (b) the SG models in the active mode. Red and blue curves are plots for $\theta=\pi/2$ (symmetric) and $\pi/4$ (asymmetric), respectively. All distributions including Fig. 1(c) of the main text are obtained from numerical simulations
	with parameter values of $l=300$, $N=200$, $k_\textrm{B}T=4$, $R=0.5$, $L=4$, $\gamma_\textrm{th}=1$, $m=1$, $M=20$, $\hat{\gamma}=-0.14$, $\omega = 0.01$, and $\zeta=0.2$. To obtain these distributions, we averaged over $8\times 10^5$ velocity data from $10^4$ trajectories in the steady state; $80$ data  are taken from each trajectory with the time interval $\Delta t = 500 > M/\gamma \simeq 300$.
	} \label{figS:velocity_distribution}
\end{figure}

\begin{figure}
\centering
\includegraphics[width=0.9\linewidth]{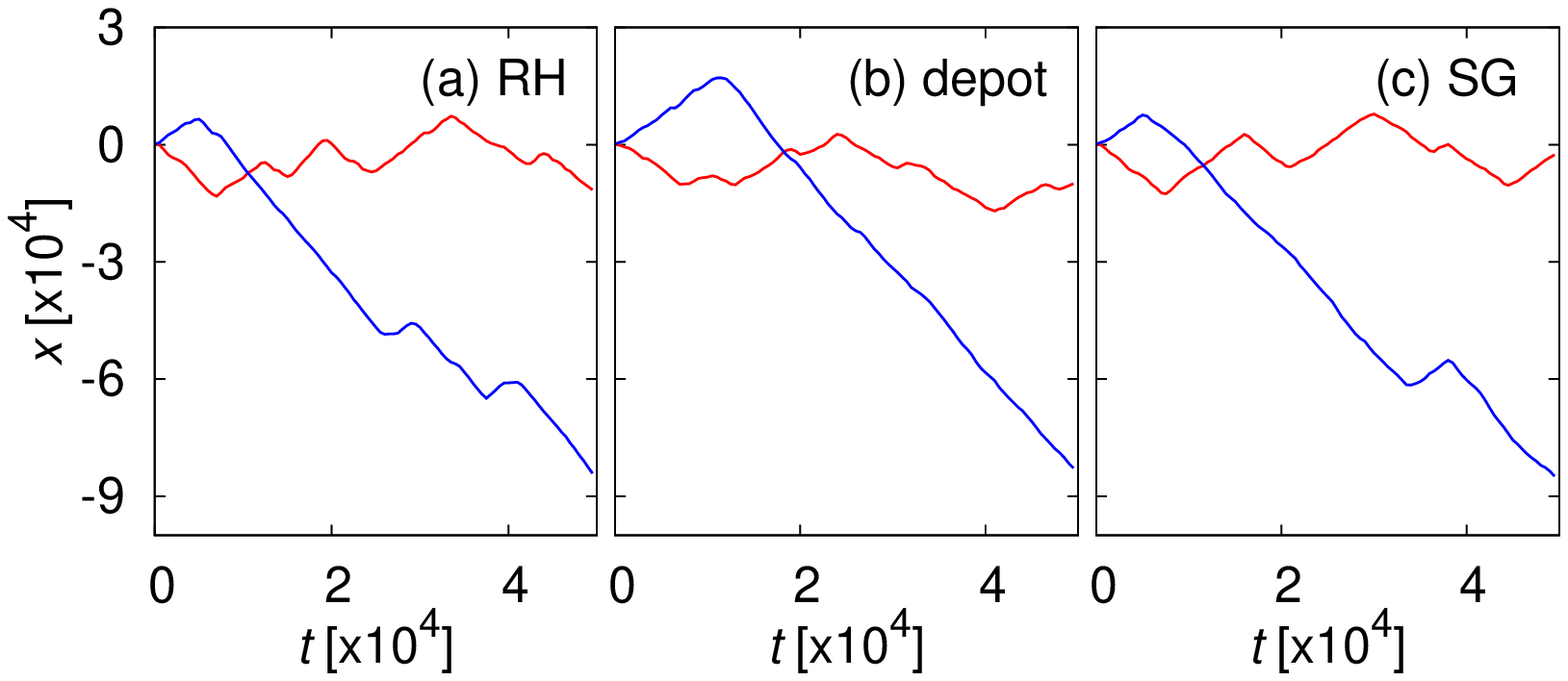}
\caption{(a), (b), and (c) are trajectories of the t-particle in the active mode of the RH, the depot, and the SG models, respectively.  Red and blue curves denote trajectories for the symmetric ($\theta = \pi/2$) and the asymmetric ($\theta = \pi/4$) case, respectively.  The symmetric-shaped particle shows a usual run-and-tumble motion with  zero-mean velocity, while the asymmetric-shaped particle shows a rectified run-and-tumble motion.
} \label{figS:active_trajectory}
\end{figure}

\begin{figure}
\centering
\includegraphics[width=0.5\linewidth]{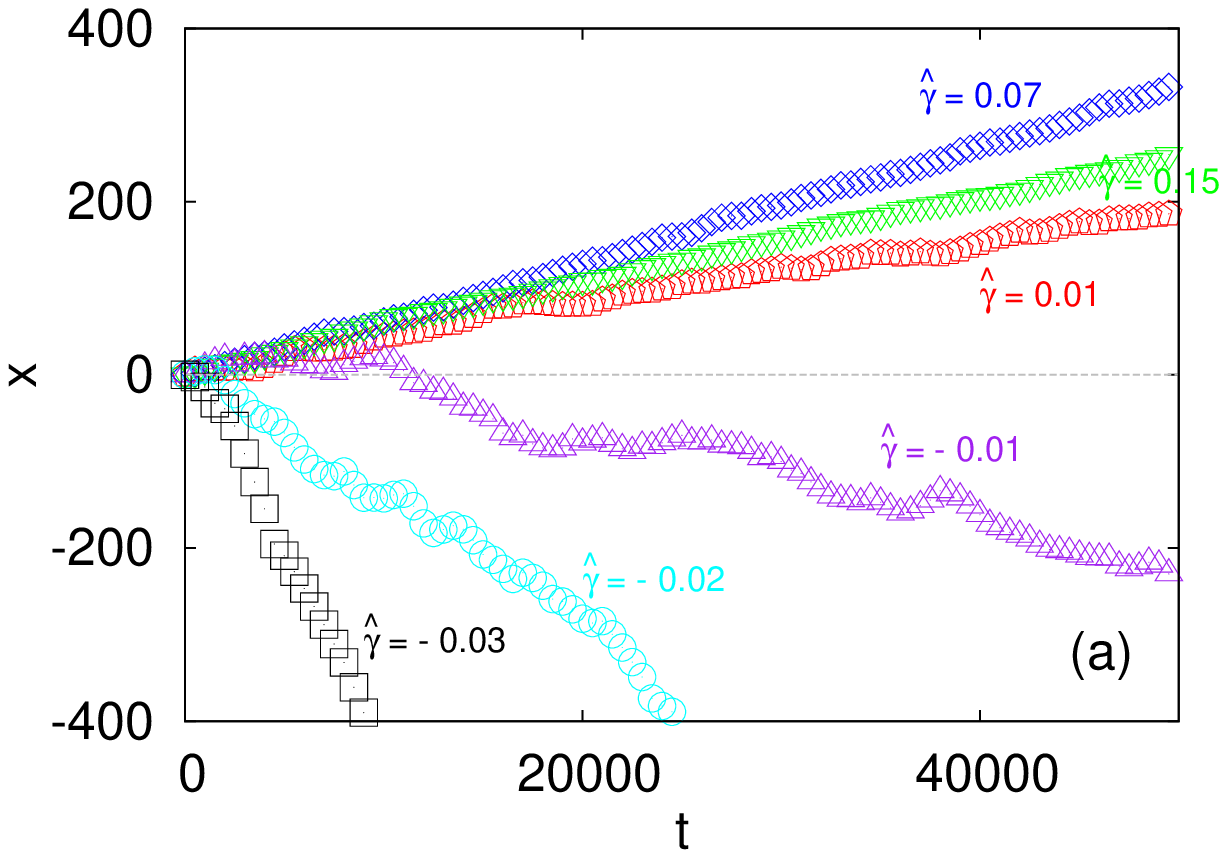}
\caption{Trajectories of the t-particle averaged over $5000$ realizations for various values of $\hat{\gamma}$ for the simple
model with $F_\textrm{drv} = -\hat{\gamma} v$.
} \label{figS:gamma_simple}
\end{figure}

With $\hat\gamma=-0.14$ and $M=20$, the numerically obtained $P_\textrm{s}(v)$ are plotted in Fig.~1(c) of the main text and in Fig.~\ref{figS:velocity_distribution}
both for the symmetric and the asymmetric cases in three different models. Typical run-and-tumble trajectories of the t-particle
are shown in Fig.~\ref{figS:active_trajectory}. The averaged trajectories for the simple model ($F_\textrm{drv}=-\hat\gamma v$)
are shown for various values of $\hat\gamma$ in Fig.~\ref{figS:gamma_simple}, which are qualitatively consistent with
Eq.~(15) of the main text.

\begin{figure}
	\centering
	\includegraphics[width=0.5\linewidth]{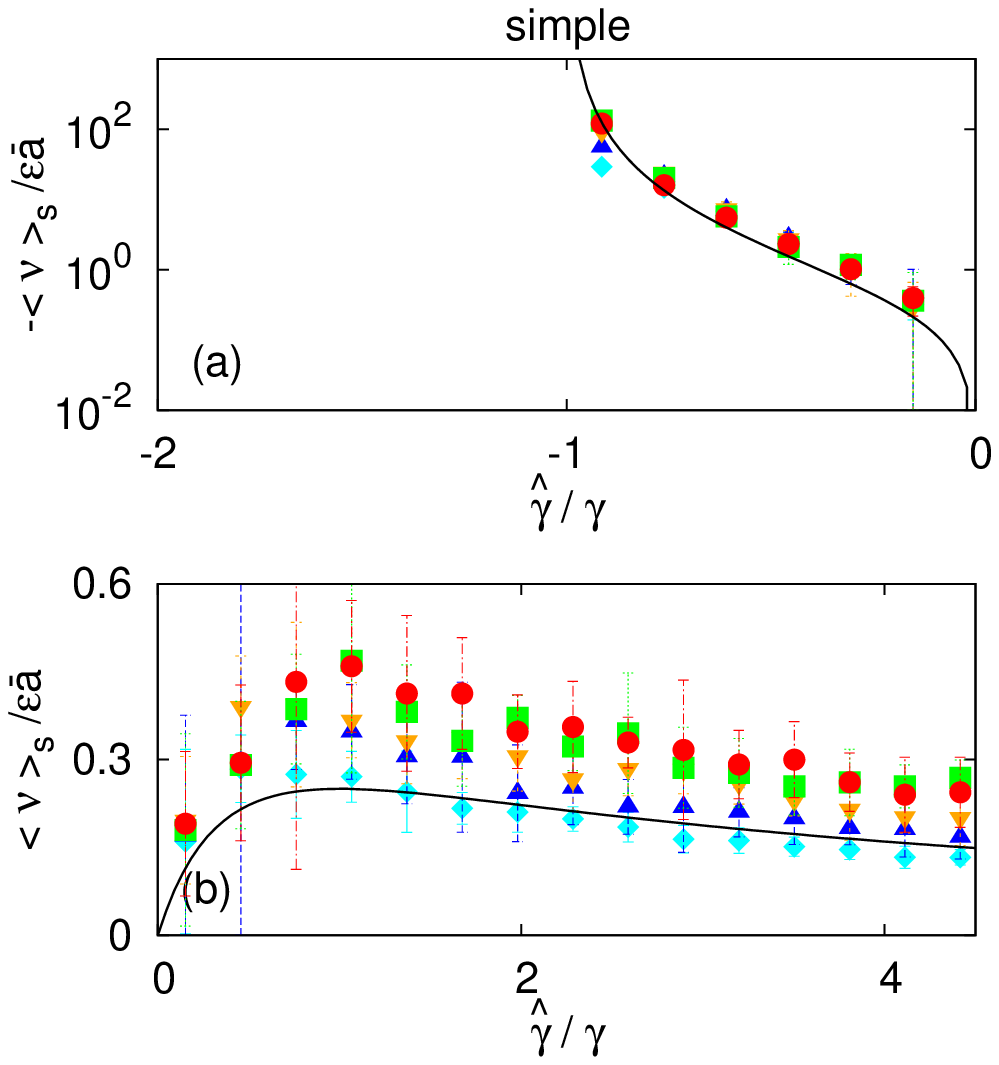}
	\caption{The scaled steady-state velocities $\langle \nu \rangle_\textrm{s} / (\bar{a} \epsilon)$ as a function of $\hat{\gamma}/\gamma$ for the simple model (a) for $\hat{\gamma}<0$ and
	(b) for $\hat{\gamma}>0$. Solid curves are analytic results. Cyan $\diamond$, blue $\vartriangle$, gold $\triangledown$,  green $\square$, red $\circ$ points denote data for $M=5$, $10$, $20$, $50$, and $100$, respectively.
	} \label{figS:v_gamma_simple}
\end{figure}

We measure the average velocities of the asymmetric t-particle for the simple, the RH, the depot, and the SG model for various $\hat\gamma$ ranging
from -0.13 to +0.29 with the t-particle mass $M=5, 10, 20, 50, 100$, which are plotted in Fig.~\ref{figS:v_gamma_simple} and Fig.~2 of the main text. For convenience, we use the scaled axes of $\hat\gamma/\gamma$ and $\langle\nu\rangle_\textrm{s}/(\bar{a}\epsilon)$.
We note that the t-particle friction coefficient $\gamma$ is obtained numerically by measuring the diffusion of the t-particle
without a driving force and using the Einstein relation  $\gamma=k_\textrm{B} T /D$ with the steady-state
position fluctuation $ \langle(\Delta x)^2\rangle_\textrm{s}=2Dt$~\cite{Risken, vanKampen}. The result is $\gamma\approx 0.067$.

\begin{figure}
	\centering
	\includegraphics[width=0.5\linewidth]{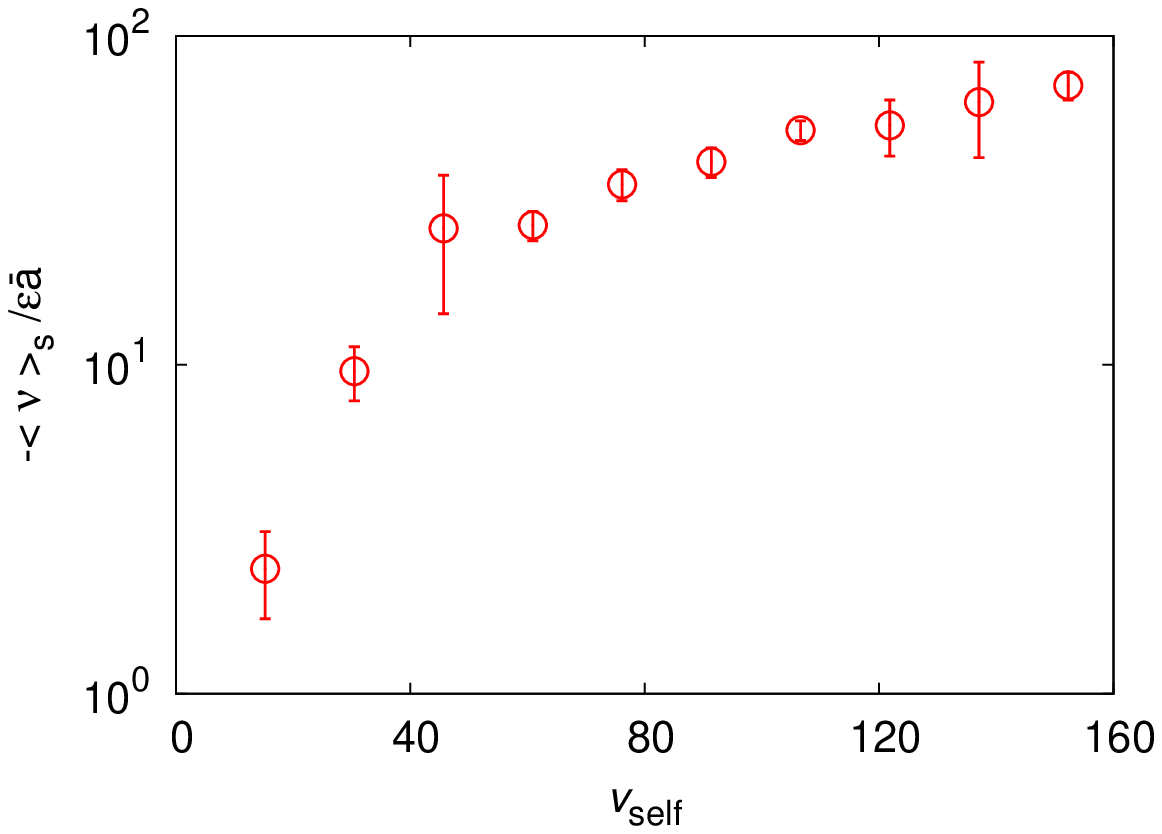}
	\caption{The scaled average velocity versus $v_\textrm{self}$ for the one-dimensional version of the active Brownian particle model with
$D_\phi=1$.
} \label{figS:gamma_speed}
\end{figure}

We also consider a driving force $F_\textrm{drv}=\gamma v_\textrm{self} \cos \phi$ with a fluctuating variable $\phi(t)$, satisfying
the dynamic equation of $\dot{\phi} = \eta$ with a Gaussian white noise $\eta(t)$ where $\langle \eta(t) \rangle =0$ and $\langle \eta(t) \eta(t^\prime)\rangle ={2 D_\phi}\delta(t-t^\prime)$. This case corresponds to a one-dimensional version of the active Brownian particle dynamics with a constant speed $v_\textrm{self}$ and diffusive angular motion~\cite{active_review, speed}. In this model, nonequilibrium activeness is given by finite $v_\textrm{self}~(\geq 0) $.
 The numerical data for the average velocity are plotted against $v_\textrm{self}$
in Fig.~\ref{figS:gamma_speed},
which are similar to the results for three other models in the active mode, i.e.~negative velocity increasing with the activity strength.

\begin{figure}
\centering
\includegraphics[width=0.5\linewidth]{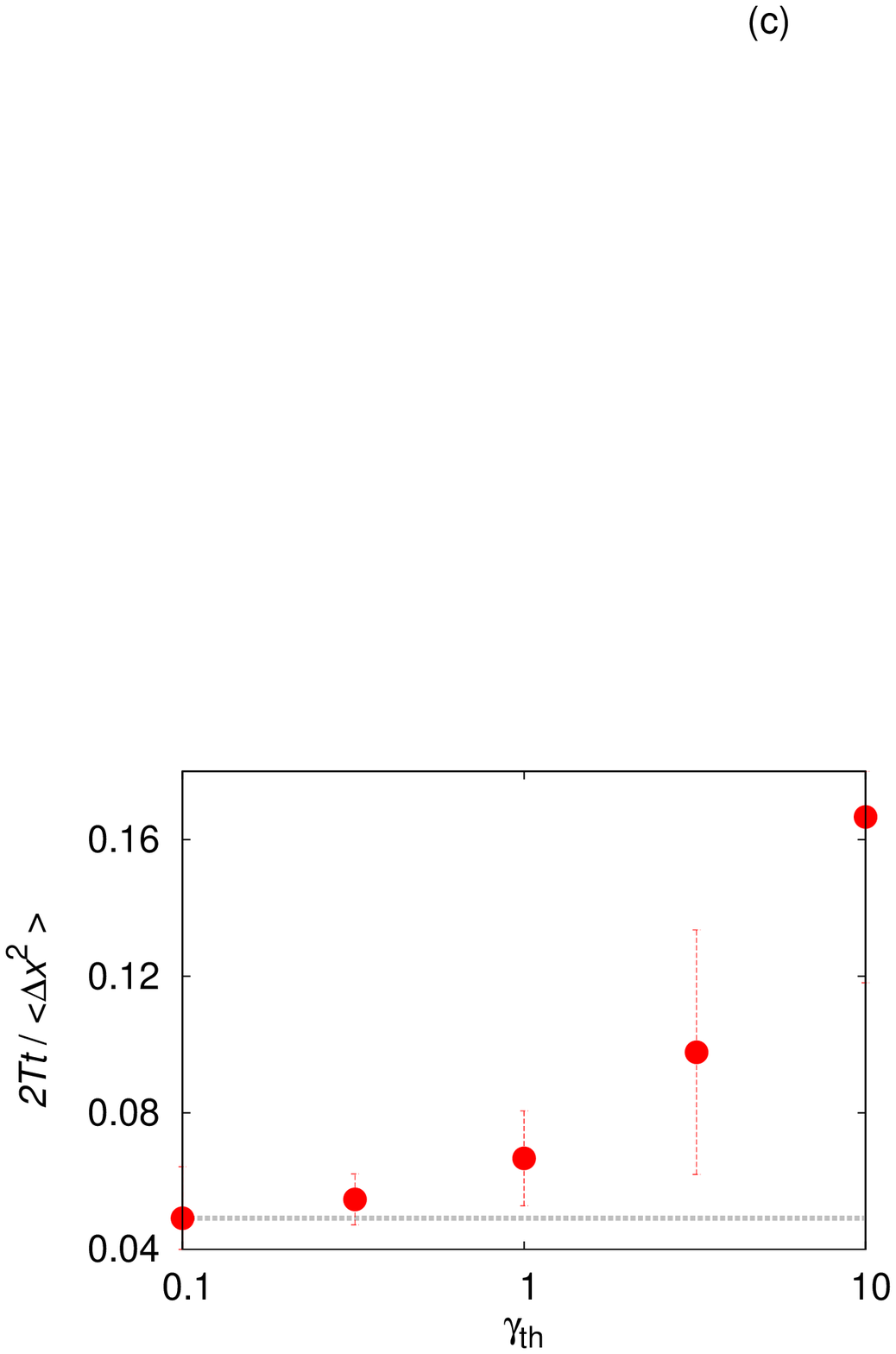}
\caption{$\gamma_\textrm{th}$-dependence of $\gamma$. The gray dotted line denotes the theoretical value of $\gamma$ directly calculated from Eq.~(6) of the main text.
} \label{figS:gamma}
\end{figure}

\emph{$\gamma_\textrm{th}$-dependence of $\gamma$} --
The friction coefficient  $\gamma$ was derived analytically in SM II
and given as
$\gamma = 4 L \rho \sqrt{ m k_\textrm{B} T  /(2\pi) } (1+\sin \theta) $ in Eq.~(6) of the main text.
Note that this result does not
depend on the thermostat dissipation coefficient $\gamma_\textrm{th}$, as it should be with an
{\em ideal} heat reservoir with a high Knudsen number, no sound wave effect, and so on. However, in molecular dynamic simulations, the reservoir effect is modeled
by kinematic collisions with a finite number of reservoir particles with a Langevin thermostat in a finite box, which
causes a spurious $\gamma_\textrm{th}$-dependence~\cite{additivity}.

Simulation results with $M=20$ and $\hat\gamma=0$ are plotted in Fig.~\ref{figS:gamma} by estimating $\gamma$ through measuring the position fluctuation and using
the Einstein relation, i.e.~$\gamma=2k_\textrm{B}Tt/\langle \Delta x^2\rangle_\textrm{s}$. It clearly depends on
and increases with $\gamma_\textrm{th}$. The theoretical value for the given set of parameters is
$\gamma\simeq 0.0484$. Interestingly, one can see that the numerical estimate approaches the theoretical value for small $\gamma_\textrm{th}$. With $\gamma_\textrm{th}=1$ (used in most of our simulations), we find $\gamma\approx 0.067$ which is $\sim 40\%$
bigger than the theoretical value. One may ascribe this overestimate
to an effectively higher r-particle density $\rho$ near the t-particle for large $\gamma_\textrm{th}$.
After colliding with the t-particle, a r-particle tends to move away but its velocity distribution relaxes back
to equilibrium quickly with large $\gamma_\textrm{th}$ (smaller relaxation time). This makes a little higher
r-particle density near the t-particle, compared to a smaller $\gamma_\textrm{th}$ case. With a higher effective $\rho$,
one can expect a bigger $\gamma$ from the above theoretical formula.

\section{Kramers-Moyal coefficients}

The Kramers-Moyal coefficients $\{a_n(v)\}$ (``jump moments'') for a single object with an arbitrary convex shape
in an ideal heat reservoir were derived in~\cite{Broeck, Meurs}.
We first show this derivation briefly here for completeness.

Consider a collision event of a r-particle with velocity ${\bf v}_r$
with the Brownian object (t-particle) with velocity $v$ on its surface point parametrized by the orientation angle
$\phi$ measured from  the horizontal ($x$) axis. After this collision, the t-particle will take the post-collision velocity $v_\phi (v,{\bf v}_r)$ determined by the given kinematics. Then, the transition rate of the t-particle velocity from $v$ to $v'$ is given as
\begin{equation}
    W(v'|v) = \int d \phi d^2 {\bf v}_r
    \delta(v' - v_\phi(v,{\bf v}_r)) w_\phi (v,{\bf v}_r) P({\bf v}_r),
    \label{eqS:transitionrate}
\end{equation}
where $P({\bf v}_r)$ is the equilibrium velocity
distribution of r-particles,
$w_\phi(v,{\bf v}_r)$ is the rate of such a collision event,
and $\delta$ is the Dirac delta function.
We assume that the r-particles are always in thermal equilibrium with temperature $T$ (ideal reservoir), regardless of collision events, leading to
the Maxwellian velocity distribution as
\begin{equation}
    P({\bf v}_r) = \frac{m}{2 \pi k_B T} \exp
    \left (- \frac{m}{2 k_B T} |{\bf v}_r|^2 \right ). \label{eqS:P}
\end{equation}

The collision rate $w_\phi(v,{\bf v}_r)$ can be obtained by the product of the
collision cross section and the incoming r-particle flux rate normal to the collision surface.
The cross section can be written as $SF_\phi$ with the total circumference of the t-particle surface $S$ and
its angular fraction $F_\phi$ (``form factor''). For the triangular shape here, we can easily find
\begin{align}\label{Ff_t_pt}
   S &= L+\frac{L}{\sin{\theta}}, \nonumber\\
    F_\phi &= \frac{L}{S} \delta(\phi - \frac{\pi}{2})
    + \frac{L}{2 S \sin{\theta}} \delta(\phi - \pi -\theta)
    + \frac{L}{2 S \sin{\theta}} \delta(\phi - 2 \pi +\theta).
\end{align}
Meanwhile, the particle flux rate is determined by the normal component of the relative velocity of two particles times the r-particle
density. Thus,
the collision rate can be written as
\begin{equation}
    w_\phi(v,{\bf v}_r) = S F_\phi~
    \rho \left |(v\hat{\bf e}_x - {\bf v}_r)\cdot\hat{\bf e}_{\phi\perp} \right |
    ~H \left ((v\hat{\bf e}_x - {\bf v}_r)\cdot\hat{\bf e}_{\phi\perp} \right)
    \label{eqS:W}
\end{equation}
with the r-particle density  $\rho$, the unit vectors $\hat{\bf e}_x$($\hat{\bf e}_y$) of the $x$($y$) axis, and the unit vector $\hat{\bf e}_{\phi\perp} = (\sin \phi) \hat{\bf e}_x - (\cos \phi) \hat{\bf e}_y$ normal to the surface. The Heaviside step function $H(x)$ guarantees a collision by excluding the outgoing flux: $H(x)=1$ for $x>0$, and $0$ for $x<0$.

The post-collision velocity $v_\phi (v,{\bf v}_r)$ of the t-particle
is determined by the elastic kinematics, such that
the kinetic energy and the $x$-component momentum are conserved,
while the $y$-component momentum is not conserved due to the constraint of the t-particle motion only in the $x$ direction.
Instead, by assuming only the normal force reacting on the colliding r-particle,
the r-particle momentum parallel to the surface is conserved. These three conservation laws read
\begin{align}
    \frac{1}{2} M v^2 + \frac{1}{2} m |{\bf v}_r|^2
    &= \frac{1}{2} M v_\phi^2 + \frac{1}{2} m |{\bf v}'_r|^2, \nonumber\\
    M v + m {\bf v}_r\cdot\hat{\bf e}_x
    &= M v_\phi + m {\bf v}'_r\cdot\hat{\bf e}_x,\\
    m {\bf v}_r\cdot\hat{\bf e}_{\phi\parallel}
    &= m {\bf v}'_r\cdot\hat{\bf e}_{\phi\parallel}, \nonumber
    \label{eqS:cons}
\end{align}
where ${\bf v}'_r$ is the post-collision velocity
of the r-particle and  $\hat{\bf e}_{\phi\parallel} = (\cos \phi) \hat{\bf e}_x + (\sin \phi) \hat{\bf e}_y$ is the unit vector
parallel to the surface with the orientation angle $\phi$.
It is straightforward to find the solution for $v_\phi$ as
\begin{equation}
    v_\phi(v,{\bf v}_r) = v + \left ( {\bf v}_r\cdot\hat{\bf e}_x
    - v - {\bf v}_r \cdot \hat{\bf e}_y \cot \phi \right)/\beta , \label{eqS:V}
\end{equation}
where $\beta = ( 1 + \epsilon^2 \sin^2 \phi ) / ( 2 \epsilon^2 \sin^2 \phi )$
with $\epsilon=\sqrt{m/M}$.

By inserting Eqs.~\eqref{eqS:P},  \eqref{eqS:W}, and \eqref{eqS:V} into Eq.~\eqref{eqS:transitionrate} and integrating it over ${\bf v}_r$, we find
\begin{equation}
    W( v+r | v ) =
    - S \rho r \int d \phi F(\phi)
    H \left ( -r \sin{\phi} \right)
    \sqrt{\frac{\alpha^2}{\pi}} \beta^2 \sin \phi
    \exp \left( - \alpha^2 \left ( v + \beta r \right )^2 \right)
\end{equation}
with the jump amplitude $r$ ($v'=v+r$) and  $\alpha = \sqrt{m / 2 k_B T} \sin \phi$.
The Kramers-Moyal coefficients $a_n (v)$ defined in Eq.~(3) of the main text
is then split into two parts
\begin{equation}
    a_n (v)
    = S \rho \sqrt{\frac{2 k_B T}{\pi m}} \left(
    - \int_{-\infty}^{0} d r \int_{\alpha > 0} d \phi
    + \int_{0}^{\infty} d r \int_{\alpha < 0} d \phi
    \right) F(\phi)
    \alpha^2 \beta^2
    r^{n+1} e^{- \alpha^2 \left ( v + \beta r \right )^2}.
\end{equation}
Using the following integral formula
\begin{equation}
\begin{split}
    &a^2 b^2 \int_0^{\infty} dr r^{n+1} e^{-a^2 (v + b r)^2} \\
    &= \frac{e^{-a^2 v^2}}{2 |a b|^{n}}
    \left[ \Gamma \left ( 1 + \frac{n}{2} \right)
    \Phi \left( 1+\frac{n}{2}, \frac{1}{2}, a^2 v^2 \right)
    - 2v \frac{|ab|}{b}  \Gamma \left ( \frac{3+n}{2} \right )
    \Phi \left ( \frac{3+n}{2},\frac{3}{2}, a^2 v^2 \right )
    \right],
\end{split}
\end{equation}
we finally obtain
\begin{equation}
\begin{split}
    a_n (v)
    = &S \rho \int d\phi F(\phi)
    \frac{1}{2} \sqrt{\frac{2 k_B T}{\pi m}} (- \alpha \beta)^{-n} e^{-\alpha^2 v^2} \\
    & \times \left[
    \Gamma \left( 1 + \frac{n}{2} \right)
    \Phi \left( 1 + \frac{n}{2}, \frac{1}{2}, \alpha^2 v^2 \right)
    + 2 \alpha v \Gamma \left( \frac{3+n}{2} \right )
    \Phi \left( \frac{3+n}{2}, \frac{3}{2}, \alpha^2 v^2 \right)
    \right]
\end{split}
\end{equation}
with the Gamma function
\begin{equation}
    \Gamma(z) = \int_0^\infty dx x^{z-1} e^{-x}
\end{equation}
and the Kummer's function
\begin{equation}
    \Phi(a,b,z) = \frac{\Gamma(b)}{\Gamma(b-a)\Gamma(a)}
    \int_0^1 dt e^{z t} t^{a-1} (1-t)^{b-a-1}.
\end{equation}

The modified Kramers-Moyal coefficients $A_n(\nu)$ defined in Eq.~(8) of the main text
becomes
\begin{align}
    A_n (\nu)
    = S \rho \left(\frac{M}{\gamma}\right) &\int d\phi F(\phi)
    \frac{1}{2} \sqrt{\frac{2 k_B T}{\pi m}} (- \eta \beta)^{-n} e^{-\eta^2 \nu^2}
    \left [
    \Gamma \left( 1 + \frac{n}{2} \right)
    \Phi \left( 1 + \frac{n}{2}, \frac{1}{2}, \eta^2 \nu^2 \right) \right . \nonumber\\
    & \left . + 2 \eta \nu \Gamma \left( \frac{3+n}{2} \right )
    \Phi \left( \frac{3+n}{2}, \frac{3}{2}, \eta^2 \nu^2 \right)
    \right ] -  G( \nu ) \nu \delta_{n,1}
\end{align}
where $\eta = v_0 \alpha = \epsilon \sin \phi/{\sqrt{2}} $.

In the small mass-ratio limit ($\epsilon \ll 1$),
$A_n(\nu)$ can be expanded in terms of $\epsilon$ as
\begin{align}\label{ex_An}
    A_n(\nu) &= S \rho \left(\frac{M}{\gamma}\right)
    (-1)^n 2^{\frac{3}{2}n} \sqrt{\frac{k_B T}{2 \pi m}}
    \Bigg [
    \Gamma \left ( 1 + \frac{n}{2} \right ) \langle \sin^n \phi \rangle_{F}~ \epsilon^n + \sqrt{2} \Gamma \left ( \frac{3+n}{2} \right )
    \langle \sin^{n+1} \phi \rangle_{F}~ \nu \epsilon^{n+1}\nonumber\\
    &+ \Gamma \left ( 1 + \frac{n}{2} \right ) \langle \sin^{n+2} \phi \rangle_{F}
    \left ( -n + \frac{1}{2} (1+n) \nu^2 \right ) \epsilon^{n+2}
    + {\cal O}(\epsilon^{n+3})
    \Bigg ]  - G( \nu ) \nu \delta_{n,1},
\end{align}
where the geometric factors are given as
\begin{equation}
    \langle \sin^n \phi \rangle_{F} \equiv\int_0^{2\pi} d\phi~ \sin^n\phi~  F_\phi=~
    \frac{L}{S} \left [ 1 - ( - \sin \theta )^{n-1} \right ]\quad (n=1,2,\ldots)
\end{equation}
with the form factor $F_\phi$ in Eq.~\eqref{Ff_t_pt}.

The friction coefficient $\gamma$ is determined from the average equation of motion for
the t-particle as
$M\langle \dot{v}\rangle=-\gamma\langle v\rangle$, which is identical to the condition on
the first Kramer-Moyal coefficient as
$a_1(v)=-(\gamma/M)v$ or equivalently $A_1(\nu)=-\nu -G(\nu)\nu$ to the lowest order in $\epsilon$.
From Eq.~\eqref{ex_An}, it is easy to identify
\begin{align}
\gamma=4 L \rho \sqrt{\frac{m k_\textrm{B} T }{2\pi} } (1+\sin \theta)~.
\end{align}
With this $\gamma$, the first few coefficients $A_n(\nu)$ can be easily obtained up to ${\cal O}(\epsilon)$ and
 given in Eq.~(9) of the main text.

\section{Analytic results for active models}

Here we present the explicit calculations for the steady-state velocity and its fluctuation
up to ${\cal O}(\epsilon)$ for the simple, the RH, the depot, and the SG models.
We use the formulae given in Eqs.~(12), (13), and (14) of the main text with $G(\nu)=\Gamma(v_0\nu)/\gamma$ for each model as
\begin{align}\label{def:func_G}
    G^\textrm{sim} (\nu) =
    \hat{\Gamma} , \quad
    G^\textrm{RH} (\nu) =
    \hat{\Gamma} + \Omega \nu^2, \quad
    G^\textrm{dpt} (\nu) =
    \hat{\Gamma} / (1 + \mathrm{Z} \nu^2), \quad
    G^\textrm{SG} (\nu) =
    \hat{\Gamma} / |\nu|,
\end{align}
with dimensionless parameters $\hat{\Gamma} = \hat{\gamma}/\gamma$,
$\Omega = \omega v_0^2 / \gamma$, and
$\mathrm{Z} = \zeta v_0^2$.

First, for the simple model, we obtain
\begin{align}
    P_\textrm{s}^{\textrm{sim}, (0)} (\nu) = \frac{1}{\mathcal{N}^\textrm{sim}} e^{- \frac{1 + \hat{\Gamma}}{2} \nu^2
    } \quad\textrm{ and} \quad g^\textrm{sim} (\nu) = 2 \hat{\Gamma} \nu - \frac{\hat{\Gamma}}{3} \left( 1 +2 \hat{\Gamma} \right )\nu^3 ,\label{eqS:sim0}
\end{align}
with the normalization factor $\mathcal{N}^\textrm{sim}=\sqrt{2\pi/(1+\hat\Gamma)}$.
Then the steady-state velocity becomes
\begin{equation}\label{eqS:simv}
    \langle \nu \rangle_\textrm{s}^\textrm{sim} \approx \bar{a} \epsilon
    \left [ 2 \hat{\Gamma} \langle \nu^2 \rangle_0^\textrm{sim}
    - \frac{\hat{\Gamma}}{3} \left( 1 +2 \hat{\Gamma} \right )\langle \nu^4 \rangle_0^\textrm{sim} \right ]
     =\bar{a} \epsilon \frac{\hat{\Gamma}}{1+\hat{\Gamma}} \langle \nu^2 \rangle_0^\textrm{sim}
    =\bar{a} \epsilon \frac{\hat{\Gamma}}{(1+\hat{\Gamma})^2} ,
\end{equation}
where we used the Gaussian integral property of $\langle \nu^4 \rangle_0^\textrm{sim}=3 (\langle \nu^2 \rangle_0^\textrm{sim})^2$ and $\langle \nu^2 \rangle_0^\textrm{sim}=\frac{1}{1+\hat\Gamma}$.
Its second moment becomes
\begin{align}
\langle \nu^2 \rangle_\textrm{s}^\textrm{sim} \approx \langle \nu^2 \rangle_0^\textrm{sim} =\frac{1}{1+\hat\Gamma}.
\end{align}

For the RH model, we obtain
\begin{align}
    &P_\textrm{s}^{\textrm{RH}, (0)} (\nu) = \frac{1}{\mathcal{N}^\textrm{RH}} e^{- \frac{1 + \hat{\Gamma}}{2} \nu^2
    - \frac{\Omega}{4} \nu^4}  \nonumber \\
 \textrm{ and}\quad   &g^\textrm{RH} (\nu) = 2 \hat{\Gamma} \nu + \left ( 2 \Omega - \frac{1}{3} \hat{\Gamma} - \frac{2}{3} \hat{\Gamma}^2 \right ) \nu^3 -\frac{1}{5} \left ( 1 + 4 \hat{\Gamma} \right ) \Omega \nu^5 - \frac{2}{7} \Omega^2 \nu^7 ,
\end{align}
with  the normalization factor $\mathcal{N}^\textrm{RH}=\int_{-\infty}^{\infty} d\nu~e^{- \frac{1 + \hat{\Gamma}}{2} \nu^2
    - \frac{\Omega}{4} \nu^4}$.
Then, the steady-state velocity becomes
\begin{equation}\label{eq:ssv_RH}
    \langle \nu \rangle_\textrm{s}^\textrm{RH} \approx \bar{a} \epsilon
    \left [ 2 \hat{\Gamma} \langle \nu^2 \rangle_0^\textrm{RH} + \left ( 2 \Omega - \frac{1}{3} \hat{\Gamma} - \frac{2}{3} \hat{\Gamma}^2 \right ) \langle \nu^4 \rangle_0^\textrm{RH} -\frac{1}{5} \left ( 1 + 4 \hat{\Gamma} \right ) \Omega \langle \nu^6 \rangle_0^\textrm{RH} - \frac{2}{7} \Omega^2 \langle \nu^8 \rangle_0^\textrm{RH} \right ].
\end{equation}
The normalization $\mathcal{N}^\textrm{RH}$ and the moments $\langle \nu^n \rangle_0^\textrm{RH}$ can be calculated numerically.
It is useful to rewrite all higher moments in terms of the second moment. Using the simple recurrence relations
as
$(n+1)\langle \nu^{n} \rangle_0^\textrm{RH} =(1+\hat\Gamma)\langle \nu^{n+2} \rangle_0^\textrm{RH} +\Omega \langle \nu^{n +4} \rangle_0^\textrm{RH}$ for $n=0,2,4\ldots$ (easily derived by the integration by parts), we find for finite $\Omega$
\begin{equation}
    \langle \nu \rangle_\textrm{s}^\textrm{RH}
     \approx\frac{\bar{a} \epsilon}{105 \Omega} \left[ 60 \Omega - 9 + 10 \hat{\Gamma} - 16 \hat{\Gamma}^2 - \left( (33 + 12\hat{\Gamma})\Omega - (9 - 10\hat{\Gamma} + 16\hat{\Gamma}^2) (1 + \hat{\Gamma}) \right) \langle \nu^2 \rangle_0^\textrm{RH} \right].
\end{equation}
Thus, the average velocity can be calculated from the numerical estimation of the zeroth-order second moment $\langle \nu^2 \rangle_0^\textrm{RH}$, and its second moment is simply given by
$\langle \nu^2 \rangle_\textrm{s}^\textrm{RH}\approx \langle \nu^2 \rangle_0^\textrm{RH}$ up to ${\cal O}(\epsilon)$.
One can show their asymptotic behaviors for large $\hat\Gamma$ as
\begin{align}
&\langle \nu \rangle_\textrm{s}^\textrm{RH}\simeq \frac{\bar{a} \epsilon}{\hat\Gamma},\quad \quad\quad\quad\quad
\langle \nu^2 \rangle_\textrm{s}^\textrm{RH}\simeq \frac{1}{\hat\Gamma} \quad \quad\textrm{for}\quad \hat\Gamma\rightarrow +\infty \nonumber\\
\textrm{ and}\quad   & \langle \nu \rangle_\textrm{s}^\textrm{RH}\simeq -{\frac{16 \bar{a} \epsilon}{105}}
\left(\frac{\hat{\Gamma}^4}{\Omega^2}\right),\quad
\langle \nu^2 \rangle_\textrm{s}^\textrm{RH}\simeq \frac{|\hat\Gamma|}{\Omega} \quad \textrm{for}\quad \hat\Gamma\rightarrow -\infty~.
\label{eqS:asymRH}
\end{align}
Note that the asymptotic behavior for positive $\hat\Gamma$ is identical to that of the simple model.

For the depot model, we obtain
\begin{align}
    &P_\textrm{s}^{\textrm{dpt}, (0)} (\nu)
    = \frac{1}{\mathcal{N}^\textrm{dpt}} e^{-\frac{1}{2} \nu^2} \left ( 1 + \mathrm{Z} \nu^2 \right )^{-\frac{\hat{\Gamma}}{2 \mathrm{Z}}}  \nonumber \\
   \textrm{ and}\quad &g^\textrm{dpt} (\nu) = - \frac{\hat{\Gamma}}{\mathrm{Z}} \nu + \frac{\hat{\Gamma} (2 \mathrm{Z} + \hat{\Gamma})}{\mathrm{Z}} \frac{\nu}{1 + \mathrm{Z} \nu^2} + \frac{\hat{\Gamma} (1 - \hat{\Gamma})}{\mathrm{Z} \sqrt{\mathrm{Z}}} \tan^{-1} \left( \sqrt{\mathrm{Z}} \nu \right), \label{eqS:depot_P_and_g}
\end{align}
with the normalization factor $\mathcal{N}^\textrm{dpt}$. Then, the steady-state velocity of the depot model becomes
\begin{equation}
     \langle \nu \rangle_\textrm{s}^\textrm{dpt} \approx \bar{a} \epsilon \left[ - \frac{\hat{\Gamma}}{\mathrm{Z}} \langle \nu^2 \rangle_0^\textrm{dpt} + \frac{\hat{\Gamma} (2 \mathrm{Z} + \hat{\Gamma})}{\mathrm{Z}}  \left \langle \frac{\nu^2}{1 + \mathrm{Z} \nu^2} \right \rangle_0^\textrm{dpt} + \frac{\hat{\Gamma} (1 - \hat{\Gamma})}{\mathrm{Z} \sqrt{\mathrm{Z}}} \left \langle \nu \tan^{-1} \left( \sqrt{\mathrm{Z}} \nu \right ) \right \rangle_0^\textrm{dpt}  \right]. \label{eqS:depot_v}
\end{equation}
 with $\langle \nu^2 \rangle_\textrm{s}^\textrm{dpt} \approx \langle \nu^2 \rangle_0^\textrm{dpt}$.
It is not possible to rewrite this expression in a simpler form, so we perform numerical integrations directly.
The asymptotic behaviors are similar to those for the RH model in Eq.~\eqref{eqS:asymRH}.

For the SG model, we obtain
\begin{align}
    P_\textrm{s}^{\textrm{SG},(0)} (\nu) = \frac{1}{\mathcal{N}^\textrm{SG}}
    e^{- \frac{1}{2} \nu^2 - \hat{\Gamma} |\nu|} \quad \textrm{and}
\quad     g^\textrm{SG} (\nu) = 2 \hat{\Gamma} \frac{\nu}{|\nu|} - 2 \hat{\Gamma}^2 \nu - \frac{1}{2} \hat{\Gamma} \nu |\nu|, \label{eqS:SG_P_and_g}
\end{align}
with the normalization factor $\mathcal{N}^\textrm{SG} =\sqrt{2 \pi} e^{\frac{1}{2}\hat{\Gamma}^2} \mathrm{erfc} \left (\frac{\hat{\Gamma}}{\sqrt{2}} \right )$ where the complimentary error function is defined as
$\mathrm{erfc}(x)=\frac{2}{\sqrt{\pi}}\int_x^\infty ds~e^{-s^2}$.
Then, the steady-state velocity becomes
\begin{equation}
    \langle \nu \rangle_\textrm{s}^\textrm{SG} = \bar{a} \epsilon
    \left ( 2 \hat{\Gamma} \langle |\nu| \rangle_0^\textrm{SG} - 2 \hat{\Gamma}^2 \langle |\nu|^2 \rangle_0^\textrm{SG} - \frac{1}{2} \hat{\Gamma} \langle |\nu|^3 \rangle_0^\textrm{SG} \right ).
\end{equation}
Again, we find useful recurrence relations as $\langle |\nu|^n \rangle_0^\textrm{SG} = (n-1) \langle |\nu|^{n-2} \rangle_0^\textrm{SG} - \hat{\Gamma} \langle |\nu|^{n-1} \rangle_0^\textrm{SG} + (2/\mathcal{N}^\textrm{SG}) \delta_{n,1}$ for $n=1,2,\ldots$ (by simple integrations by parts). Then, we obtain a simplified form for $\langle \nu \rangle_\textrm{s}^\textrm{SG}$ as
\begin{align}
    \langle \nu \rangle_\textrm{s}^\textrm{SG}
    = \frac{\bar{a} \epsilon \hat{\Gamma}}{2} \left ( -3 \hat{\Gamma} + (2 + 3 \hat{\Gamma}^2) \langle |\nu| \rangle_0^\textrm{SG} \right )
    = \frac{\bar{a} \epsilon \hat{\Gamma}}{2} \left (
    - 5 \hat{\Gamma} - 3 \hat{\Gamma}^3 + \frac{4 + 6 \hat{\Gamma}^2 }{\mathcal{N}^\textrm{SG}}
    \right ).
\end{align}
with the second moment
\begin{align}
\langle \nu^2 \rangle_\textrm{s}^\textrm{SG} \approx \langle |\nu|^2 \rangle_0^\textrm{SG}=
1+\hat{\Gamma}^2-\frac{2\hat\Gamma}{{\mathcal{N}^\textrm{SG}}}.
\end{align}
Their asymptotic behaviors are given as
\begin{align}
&\langle \nu \rangle_\textrm{s}^\textrm{SG}\simeq   \bar{a} \epsilon
\left(-2+\frac{13}{\hat{\Gamma}^2}\right),~
\langle \nu^2 \rangle_\textrm{s}^\textrm{SG}\simeq \frac{2}{\hat{\Gamma}^2} \quad \textrm{for}\quad \hat\Gamma\rightarrow +\infty \nonumber\\
\textrm{ and}\quad   & \langle \nu \rangle_\textrm{s}^\textrm{SG}\simeq - {\frac{3\bar{a} \epsilon}{2}\hat{\Gamma}^4},\quad\quad\quad
\langle \nu^2 \rangle_\textrm{s}^\textrm{SG}\simeq \hat{\Gamma}^2 \quad \textrm{for}\quad \hat\Gamma\rightarrow -\infty .
\label{eqS:asymSG}
\end{align}

\begin{figure}
	\centering
	\includegraphics[width=0.5\linewidth]{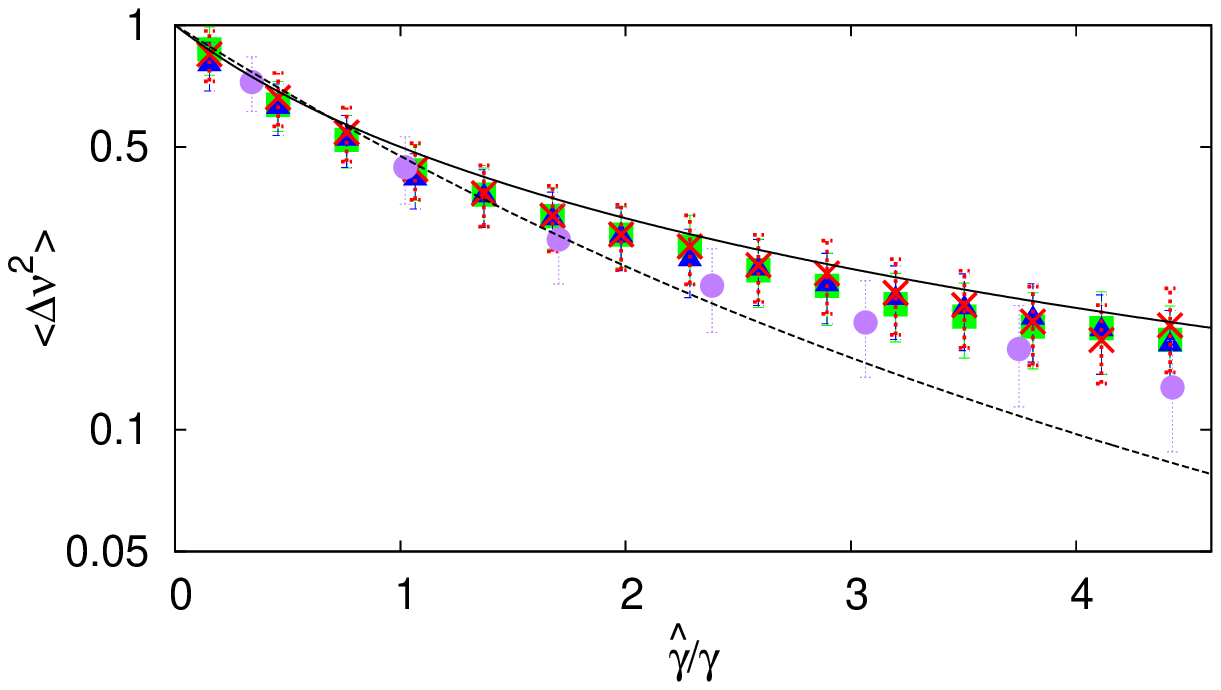}
	\caption{Fluctuations as a function of $\hat{\gamma}/\gamma$. Solid and dashed curves correspond to the simple and the SG models 
in Eq.~(S22) and Eq.~(S32), respectively. Green $\blacksquare$, blue $\blacktriangle$, red $\times$, and purple $\bullet$ points 
denote numerical data for the simple, the RH, the depot, and the SG models, respectively. } \label{figS:vel_variance}
\end{figure}


\vfil\eject